\documentclass[aps,twocolumn,showpacs,superscriptaddress]{revtex4}

\usepackage{graphicx}
\usepackage{latexsym}

\usepackage{color}
\def\Vec#1{\mbox{\boldmath $#1$}}

\begin{document}

\title{Valence bond supersolid in a bilayer extended Bose Hubbard model}

\author{Kwai-Kong Ng}
\email[]{kkng@thu.edu.tw}
\affiliation{Department of Applied Physics, Tunghai University, Taichung, Taiwan}

\def\Vec#1{\mbox{\boldmath $#1$}}
\def\degC{\kern-.2em\r{}\kern-.3em C}

\def\SimIneA{\hspace{0.3em}\raisebox{0.4ex}{$<$}\hspace{-0.75em}\raisebox{-.7ex}{$\sim$}\hspace{0.3em}} 

\def\SimIneB{\hspace{0.3em}\raisebox{0.4ex}{$>$}\hspace{-0.75em}\raisebox{-.7ex}{$\sim$}\hspace{0.3em}}

\date{\today}

\begin{abstract}
The hardcore extended Bose Hubbard model on a bilayer square lattice with attractive (repulsive) interplane (intraplane) interactions has been investigated numerically. 
Focusing on the strong interplane  hopping parameter regime, triplet valence bonds of dimers are stabilized to form valence bond checkerboard solid at quarter filling. Increasing the particle number we confirm, for the first time, the presence of the exotic valence bond supersolid phase, where the valence bond solid ordering and the boson superfluidity coexist. Rising further the particle number will lead to a checkerboard solid of dimer pair at half filling for strong intraplane repulsion, or a valence bond Mott insulator for weak repulsive interactions. We analyze the rich ground state phase diagrams of this model, which can be experimentally realized in optical lattices of cold atoms. 

\end{abstract}


\maketitle

A special commensurate valence bond supersolids (VSS) at quarter boson filling has been proposed previously on the checkerboard lattice of bosonic hardcore Hubbard model, which is studied by using Green function quantum Monte Carlo (GFQMC) techniques \cite{Ralko}. In this frustrated system, the valence bond solid phase is characterized by the mixed columnar and plaquette crystal orders, in which ways the bosons are delocalized. The  valence bond  solid phase is found to extend into the superfluid (SF) phase in a small region at the commensurate filling. A later large-scale quantum Monte Carlo (QMC) based on stochastic series expansion (SSE) method, however, found that the VSS is unstable in the thermodynamic limit \cite{Wessel}. Instead, the crystal melts through a direct transition into the SF phase. On the other hand, valence bond solid with delocalized bosons around a subset of hexagons has also been observed in Kagome lattice  of hardcore bosons \cite{Isakov}. It is demonstrated that the quantum phase transition between VBS and SF is weakly first-order and there is no intermediate VSS phase.  VSS has attracted considerable interests as it is characterized by the coherent quantum dynamics of both solid ordering and superfluid ordering, in contrast to the usual supersolids (SS) of classical solid ordering. The evidence of the presence of the VSS phase is, nevertheless, elusive. 

\begin{figure}
\includegraphics[clip,width=0.9\columnwidth]{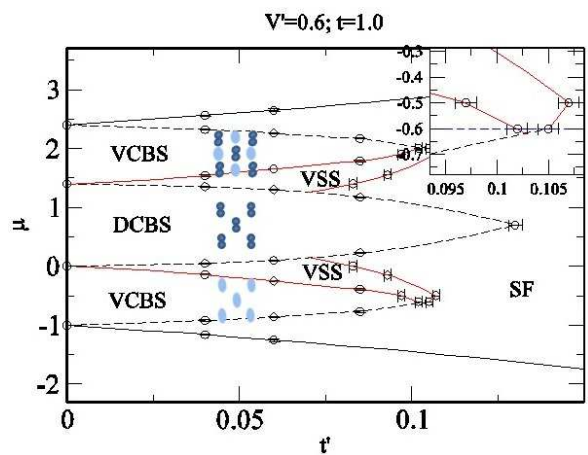}
\caption{(color online). Ground state phase diagram chemical potential $\mu$ vs intraplane hoppings $t'$ for intraplane repulsion $V'=0.6$. Valence bond supersolid (VSS) is found in between the half-filled dimer checkerboard solid (DCBS) and quarter-filled valence bond checkerboard solid (VCBS). The checkerboard structures of the phases are illustrated in the diagram with the ellipses represent valence bonds $|t_0\rangle$ and double circles the dimer states $|11\rangle$. The solid (dotted) lines denote second (first) order phase transitions. Lattice size $L=24$ and temperature $T=0.005$ are used for all calculations. The inset shows the enlarged region around the tip of the lobe at $\mu=-0.6$.}
\label{fig1}
\end{figure}

Employing the QMC approach, in this work we study numerically the coupled bilayer system of extended Bose-Hubbard model with attractive interplane interactions.  To allow the possibility of broken crystal symmetry, intraplane repulsive interaction $V'$ is introduced. The ground state phase diagram is presented in Fig. \ref{fig1}. For large interplane hopping $t$, the triplet state $|t_0\rangle=\frac{1}{\sqrt{2}}(|10 \rangle + |01 \rangle)$ ($|n_1 n_2 \rangle$ denotes the dimer state with $n_1$ ($n_2$) boson on the layer 1 (2).) is energetically favored and this unusual symmetric $|t_0\rangle$ valance bond state is a counterpart to the more common valence bond of antisymmetric singlet state. 
At quarter filling the ground state is a checkerboard solid of these valence bonds (VCBS). Doping bosons on the VCBS, we find for the first time a thermodynamically stable VSS phase, characterized by the coexistence of valence bond checkerboard order and the superfluidity of bosons.  By finite size scaling, we observe that the VCBS phase changes to VSS through a 3D XY continuous quantum phase transition and the VSS melts via a 3D Ising-type continuous transition to a superfluid as indicated by the red lines in Fig. \ref{fig1}. Furthermore, the half filling ground states can be a dimer checkerboard solid (DCBS) or valence bond Mott insulator (VMI) \cite{Block,Carrasquilla} depends on the intraplane repulsions $V'$ (see Fig. \ref{fig4}). 

It is noted that considerable efforts has been made in related bilayer systems of attractive interplane coupling \cite{Trefzger,Safavi,Macia}, owing to  the experimental breakthroughs in the ultracold gases of atoms or molecules \cite{Lewenstein} that allow the realizations and investigations of these many-body quantum systems. Condensations of composite bosons, like pair superfluids and pair supersolids \cite{Trefzger,Safavi,Macia}, are resulted from the strong interplane attractions. Hoppings between different layers are usually ignored in order to lower the interplane kinetic energy to preserve the pairs. The competitions between the interplane kinetic energy and interaction energy, however, may drive to unexpected quantum ground states. 

The model under discussion is a system of interacting hardcore bosons confined in a bilayer of two dimensional square lattices, of which the Hamiltonian is given by

\begin{eqnarray}
H=&-&t \sum_i (b_{1,i} ^\dagger b_{2,i}+ h.c.) + V \sum_i  n_{1,i} n_{2,i}  \\ 
&-& t' \sum_{\alpha,\langle i,j\rangle}  (b^\dagger_{\alpha,i} b_{\alpha,j} + h.c.) \nonumber \\ 
&+& V'  \sum_{\alpha,\langle i,j\rangle} n_{\alpha,i} n_{\alpha,j} -\mu \sum_{\alpha,i} n_{\alpha,i}. \nonumber
\label{Ham}
\end{eqnarray}

\noindent where $t (t')$ and $V (V')$ are the interlayer (intralayer) hopping and interaction terms respectively. $b_{\alpha,i}$($b_{\alpha,i} ^\dagger$) are the boson annihilation (creation) operators at the site $i$ of layer $\alpha$ with $n_{\alpha,i}$  the corresponding number operators.  In the grand canonical ensemble, the total boson number is controlled by a single chemical potential $\mu$. We consider in this work the attractive interlayer interaction and set $V=-1$ as the energy scale. We will also focus on the strong interplane hopping regime that $t=1$.

This Hamiltonian is similar to that of the two species of interacting bosons \cite{Altman,Isacsson,Kuklov01,Kuklov02,Hu,Iskin,Li}, or the three body constrained bosons in a single layer \cite{Daley,Lee,Bonnes,Ng,Singh}, which have been intensively discussed recently. In these systems the bosons form pairs, or even multimers in some long range dipolar bosonic systems. Boson pairs may condense to form pair superfluids or pair supersolids depend on the interplay among the interaction energy and the kinetic energy of the dimers. However, what distinguished the current Hamiltonian $H$ from the previous works is the introduction of interplane hopping $t$, which is equivalent to an exchange process of different types of bosons in the multi-species systems.  This interplane hopping competes with the interplane attraction and leads to the stabilization of quantum states such as valence bond solids and supersolids. In the spin language, $H$ is simply a system of two antiferromagnetic layers that ferromagnetically coupled together. Possibly due to the absence of related magnetic materials, these ferromagnetic coupled  systems did not attract many attentions before. Nonetheless, these interactions can now be realized in cold atoms on optical lattices. 

The ground states of independent dimers are obvious: for dominant attractive interactions $V$, pair of occupied boson $|11 \rangle$ is favored whereas for dominant interplane hoppings $t$ the valence bond $|t_0\rangle$ is favored. Turning on the interdimer repulsions and hoppings leads to a variety of quantum ground states, including paired SF, VMI, and VBSS. Our investigation here focuses on the dominant $t$.

The numerical method employed in this work is based on the well-established quantum Monte Carlo approach of the  SSE algorithm \cite{Sandvik}. 
In the simulation, the broken symmetries of the different quantum states in the system are identified by the following order parameters. The diagonal long-range order of solid phases can be identified by the structure factor 

\begin{eqnarray}
S(\mathbf{k})=\frac{1}{N^2}\sum_{i,j}\left\langle n_i n_j \right\rangle e^{ i \mathbf{k}\cdot \left( \mathbf{r}_i - \mathbf{r}_j \right)  }, \nonumber
\label{S(k)}
\end{eqnarray}

\noindent where $n_i=\sum_\alpha n_{\alpha,i}$. In particular, the dimer checkerboard crystal structure is signaled by $S(\pi,\pi)$. 
Likewise the VCBS is characterized by the valence bond structure factor $S_v(\pi,\pi)$ with $S_v(\mathbf{k})$ \cite{Isakov} defined as:

\begin{eqnarray}
S_v(\mathbf{k}) &=& \frac{1}{N'^2}\sum_{i,j} \left\langle B_i B_j \right\rangle e^{i \mathbf{k} \cdot \left( \mathbf{r}_i-\mathbf{r}_j \right) }. \nonumber \\
B_i &=& t(b_{1,i} ^\dagger b_{2,,i}+ b_{1,i} b_{2,i} ^\dagger) 
\label{S_V(k)}
\end{eqnarray}

\noindent with $N'=N/2$, number of sites on a single layer.  For the superfluid phase, the convenient order parameter used in SSE is the superfluidity $\rho_s=(\left\langle W^2_x + W^2_y \right\rangle )/(2 \beta t')$, measured through the fluctuations of winding number ($W_x$ and $W_y$) in both directions.  

\begin{figure}
\includegraphics[clip,width=0.9\columnwidth]{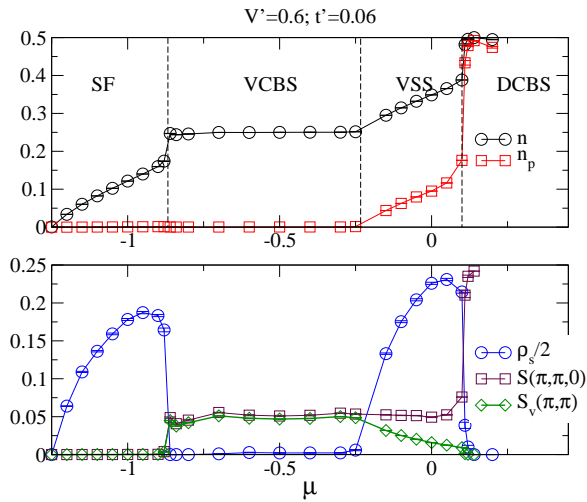}
\caption{(color online). Order parameters as a function of chemical potential $\mu$. Upper panel shows the average boson density $n$ and average pair density $n_p=\left\langle \sum_i n_{1,i}n_{2,i}/N' \right\rangle $. Lower panel shows the superfluidity $\rho_s$, boson structure factor $S(\pi,\pi)$ and valence bond structure factor $S_v(\pi,\pi)$.}
\label{fig2}
\end{figure}

Fig. \ref{fig1} shows the ground state phase diagram for intraplane repulsion $V'=0.6$ such that $V'>t-|V|/2$, which corresponds to the strong interacting pairs regime. In this regime, at half filling the bosons pairing up due to the interplane attraction but the dimers repel each other to reduce the intraplane repulsive energy. Consequently a DCBS is stabilized with a finite peak of the boson structure factor $S(\mathbf{Q})$ at $\mathbf{Q}=(\pi,\pi)$. The plateau of boson density $n$ in Fig. \ref{fig2} indicates the incompressibility of the DCBS phase. At quarter filling, on the other hand, the ground state is a checkerboard solid of valence bond. Owing to the large interplane hoppings $t$, bosons form valence bonds $|t_0\rangle$ at each rungs to avoid the pairing of bosons in spite of the attractive interaction. It is noted that the valence bond, on its own, does not break any crystal structure of the system, dissimilar to other valence bond solids \cite{Ralko,Isakov} where the valence bonds are responsible to the breaking of the crystal structure. The broken translational symmetry here is, on the other hand, introduced by the checkerboard ordering, arises from the intraplane repulsion, of the valence bonds and is identified by the finite value of bond structure factor $S_v(\pi,\pi)$ (Fig. \ref{fig2}). Further reducing the boson numbers, via a first order quantum phase transition the VCBS phase changes to SF phase as shown in the phase diagram. 

Most importantly, in between the quarter filling (or three quarter filling) and half filling, there is a valence bond supersolid phase, in which the checkerboard ordering of valence bonds and superfluidity of bosons coexist, as indicated by the finite $S_v(\pi,\pi)$ and $\rho_s$ in Fig. \ref{fig2}. Different from the VSS suggested by Ref. \cite{Ralko}, the VSS in this bilayer model does not restrict to the commensurate quarter filling but extends in a wide range of boson density from 1/4 to 1/2 filling. 
Increasing boson numbers from quarter-filled VCBS, the bond structure factor $S_v(\pi,\pi)$ reduces as the superfluidity of bosons increases (Fig. \ref{fig2}). However, it is noted that the boson structure factor $S(\pi,\pi)$ remains almost unchanged. This observation can be explained by the fact that the pair dimer density increases rapidly as fast as the boson density, which implies that the extra bosons occupied on both empty dimers and  the valence bond dimers as well. Therefore their contributions on the boson structure factor $S(\pi,\pi)$ cancels out and leaves $S(\pi,\pi)$ about the same. Nevertheless the situation for large $t'$ (not shown here) is different as the pair dimer density increases rather slowly and implies that the extra bosons are likely to occupy the empty dimers so that $S_v(\pi,\pi)$ and $S(\pi,\pi)$ reduce in about the same rate. We have also carried out finite size analysis of lattice size up to $L=36$ to verify that the VSS survives in the thermodynamical limit (see below).

We also remark that this VSS is distinct from the supersolid found in the bilayer XXZ model \cite{Ng02}, in which case the interplane interactions are repulsive and the solid ordering is characterized by the finite boson structure factor at wave vector $\mathbf{Q}=(\pi,\pi,\pi)$. On the other hand, the solid ordering VSS here is identified by a finite bond structure factor $S_v(\pi,\pi)$, as well as a finite $S(\pi,\pi)$, but with vanished  structure factor at $\mathbf{Q}=(\pi,\pi,\pi)$. In the bilayer XXZ model, the solid ordering is a direct consequence of repulsive (antiferromagnetic) interactions. On the contrary, the VSS in the present model results from the strong interplane hoppings $t$ that leads to valence bonds, and the intraplane repulsion $V'$ that leads to checkerboard ordering.

\begin{figure}\includegraphics[clip,width=0.9\columnwidth]{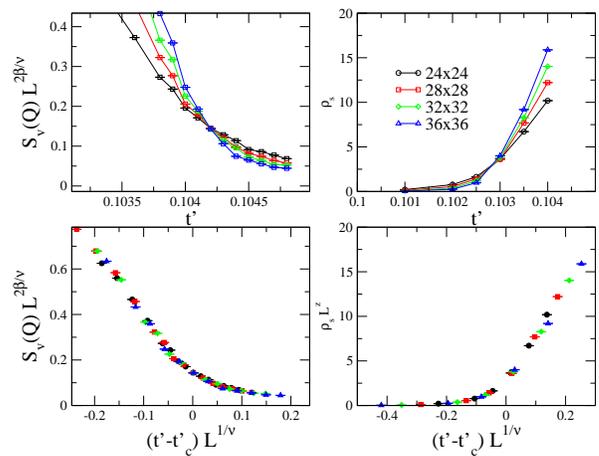}
\caption{(color online). Finite size analysis for the VCBS-VSS and VSS-SF quantum phase transitions near the tip of the quarter-filled lobe in Fig. \ref{fig1} with $\mu=-0.6$. The data collapse in the lower panels illustrates that the transitions agree with the universality class of 3D Ising type  and 3D XY type respectively.}
\label{fig3}
\end{figure}

To study the critical behavior close to the tip of the lobe of VCBS we perform a finite size analysis at $\mu=-0.6$ (along the dotted line in the inset of Fig. \ref{fig1}) as shown in Fig. \ref{fig3}. The bond  structure factor $S_v$ scales as $L^{2\beta/\nu} G((t'- t'_{c1})L^{1/\nu}, \beta/L^z)$ and the superfluidity $\rho_s$ scales as $L^{-z}F( (t'-t'_{c2})L^{1/\nu},\beta /L^z)$. The dynamical exponent $z$ is taken to be 1.
Our analysis indicates that the  $S_v(\pi,\pi)$ vanishes at $t'_{c1}=0.104194(7)$ and $\rho_s$ emerges at $t'_{c2}=0.1287(13)$ in the thermodynamical limit. A clear data collapse of $S_v(\pi,\pi)$ for different lattice sizes is observed  when the critical exponents $2\beta/\nu=1.0366$ and $\nu=0.63$ of the 3D Ising model is applied \cite{Hasenbusch}. Similarly, the continuous phase transition of $\rho_s$ is found to be consistent with the 3D XY model with critical exponent $\nu=0.662$. The finite scale analysis confirms the continuous quantum phase transitions for both quantities and there is a small region of VSS phase with quarter-filling near the tip of the lobe of the VCBS. 

\begin{figure}
\includegraphics[clip,width=0.75\columnwidth]{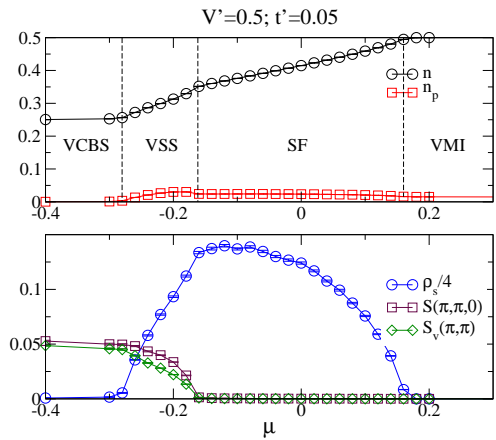}
\caption{(color online). VSS-SF and SF-VMI transition is shown for a weaker repulsion $V'=0.5$. Pair density $n_p$ is almost vanished for valence bond dominated regimes.}
\label{fig4}
\end{figure}
    
\begin{figure}
\includegraphics[clip,width=0.75\columnwidth]{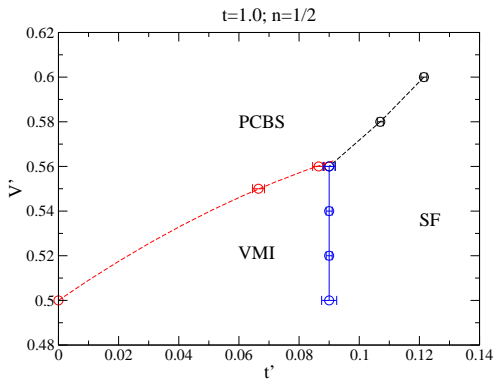}
\caption{(color online). Ground state phase diagram $V'$ vs $t'$ demonstrates the competition between VMI and PCBS at half-filling.}
\label{fig5}
\end{figure}

Another interesting feature at half filling is the presence of valence bond Mott insulator (VMI) when the repulsive interaction $V'$ is reduced to 0.5 as shown in the Fig. \ref{fig4}. When the repulsive energy is not sufficient to sustain the checkerboard structure of the DCBS, bosons rearrange themselves into dimers of valence bonds to reduce the kinetic energy cost. This is a gapped insulating phase that all dimers are occupied by valence bonds. No lattice symmetry is broken in this case and therefore the transition from SF to VMI is expected to be continuous, differs from that of SF-DCBS as discussed above.  Still, a VSS phase is stabilized in between the VCBS and SF phases. 
Keeping the  chemical potential $\mu=2V'-|V|/2$ so that the average boson density $n=1/2$, Fig \ref{fig5} shows how the gapped phases, DCBS and VMI, compete with each other as the intralayer repulsion $V'$ changes. Since the checkerboard ordering in DCBS keeps the neighboring pairs away from each other, it maintains a lower repulsive energy than the VMI phase for large $V'$. Contrastingly, the disturbance of boson hoppings on the checkerboard ordering  will cost larger repulsive energy. For VMI, however, bosons are moving in a background that breaks no translational symmetry and is more favorable for large $t'$. The first order phase boundary between the DCBS and VMI is obtained by comparing their ground state energies. For even larger $t'$ both DCBS and VMI phases melt into a SF phase via a first order and second order phase transitions respectively. 
  
To conclude, a hardcore bosonic bilayer model with interplane attractive interactions is explored. In the strong interplane hopping regime, exotic supersolid of triplet valence bonds with finite superfluidity and valence bond checkerboard ordering  is observed. The nature of the quantum phase transitions of the VSS is investigated. Besides the VSS, Mott insulator of valence bond competes with dimer checkerboard solid and is energetically favored at half filling for small $V'<t-|V|/2$. It is interesting to note that pair superfluid is stable for weak interplane hopping or for strong interplane attraction. The nature of the transitions from PSF to VCBS or VSS is of great interest and will be discussed elsewhere. It is also not surprising to find a pair supersolid phase if long-range repulsive interactions is introduced. Because the bilayer interactions can be easily realized in optical lattice, the rich phase diagrams and the related phases may be experimentally verified in the near future.

\acknowledgments
We are grateful to Min-Fong Yang for many stimulating discussions. This work was supported by NCTS,  MOST 100-2112-M-029-003-MY3 and 103-2112-M-029-003.

\end{document}